\documentclass[aps,prl,showpacs,groupedaddress,twocolumn]{revtex4}

\begin{document}
\topmargin-1.0cm


\def\bea{\begin{eqnarray}}
\def\eea{\end{eqnarray}}
\def\ben{\begin{equation}}
\def\een{\end{equation}}
\def\benu{\begin{enumerate}}
\def\enu{\end{enumerate}}

\def\n{n}

\def\sss{\scriptscriptstyle\rm}

\def\g{_\gamma}

\def\l{^\lambda}
\def\lfc{^{\lambda=1}}
\def\lo{^{\lambda=0}}

\def\marnote#1{\marginpar{\tiny #1}}
\def\rsav{\langle r_s \rangle}
\def\invdif{\frac{1}{|\br_1 - \br_2|}}

\def\hatT{{\hat T}}
\def\hatV{{\hat V}}
\def\hatH{{\hat H}}
\def\1var{(\bx_1...\bx\N)}

\def\half{\frac{1}{2}}
\def\quart{\frac{1}{4}}

\def\bp{{\bf p}}
\def\br{{\bf r}}
\def\bR{{\bf R}}
\def\bu{{\bf u}}
\def\bmu{{\bf \mu}}
\def\bx{{\br t}}
\def\by{{y}}
\def\ba{{\bf a}}
\def\bq{{\bf q}}
\def\bj{{\bf j}}
\def\bX{{\bf X}}
\def\bF{{\bf F}}
\def\bchi{{\bf \chi}}
\def\bof{{\bf f}}

\def\cA{{\cal A}}
\def\cB{{\cal B}}

\def\x{_{\sss X}}
\def\c{_{\sss C}}
\def\s{_{\sss S}}
\def\xc{_{\sss XC}}
\def\Hx{_{\sss HX}}
\def\Hc{_{\sss Hc}}
\def\Hxc{_{\sss HXC}}
\def\xj{_{{\sss X},j}}
\def\xcj{_{{\sss XC},j}}
\def\N{_{\sss N}}
\def\H{_{\sss H}}

\def\ext{_{\rm ext}}
\def\pot{^{\rm pot}}
\def\hyb{^{\rm hyb}}
\def\HF{^{\rm HF}}
\def\hah{^{1/2\& 1/2}}
\def\loc{^{\rm loc}}
\def\LSD{^{\rm LSD}}
\def\LDA{^{\rm LDA}}
\def\GEA{^{\rm GEA}}
\def\GGA{^{\rm GGA}}
\def\SPL{^{\rm SPL}}
\def\sce{^{\rm SCE}}
\def\PBE{^{\rm PBE}}
\def\DFA{^{\rm DFA}}
\def\TF{^{\rm TF}}
\def\VW{^{\rm VW}}
\def\helm{^{\rm unamb}}
\def\una{^{\rm unamb}}
\def\ion{^{\rm ion}}
\def\HOMO{^{\rm HOMO}}
\def\gs{^{\rm gs}}
\def\dyn{^{\rm dyn}}
\def\adia{^{\rm adia}}
\def\I{^{\rm I}}
\def\pot{^{\rm pot}}
\def\sav{^{\rm sph. av.}}
\def\syv{^{\rm sys. av.}}
\def\pnav{^{\rm sym}}
\def\av#1{\langle #1 \rangle}
\def\unif{^{\rm unif}}
\def\LSD{^{\rm LSD}}
\def\ee{_{\rm ee}}
\def\vir{^{\rm vir}}
\def\ALDA{^{\rm ALDA}}
\def\VUC{^{\rm VUC}}
\def\PGG{^{\rm PGG}}
\def\GK{^{\rm GK}}
\def\atom{^{\rm atmiz}}
\def\trans{^{\rm trans}}
\def\SPA{^{\rm SPA}}
\def\SMA{^{\rm SMA}}

\def\sav{^{\rm sph. av.}}
\def\syv{^{\rm sys. av.}}

\def\up{_\alpha}
\def\dn{_\beta}
\def\up{_\uparrow}
\def\dn{_\downarrow}

\def\td{time-dependent~}
\def\KS{Kohn-Sham~}
\def\DFT{density functional theory~}

\def\fourint{ \int_{t_0}^{t_1} \! dt \int \! d^3r\ }
\def\fourintp{ \int_{t_0}^{t_1} \! dt' \int \! d^3r'\ }
\def\intx{\int\!d^4x}
\def\sph_int{ {\int d^3 r}}
\def\radint{ \int_0^\infty dr\ 4\pi r^2\ }

\def\PRA{Phys. Rev. A\ }
\def\PRB{Phys. Rev. B\ }
\def\PRL{Phys. Rev. Letts.\ }
\def\JCP{J. Chem. Phys.\ }
\def\JPCA{J. Phys. Chem. A\ }
\def\IJQC{Int. J. Quant. Chem.\ }

\title {Density functional theory of dissipative systems}
\author {Kieron Burke}
\affiliation {Department of Chemistry and Chemical Biology, Rutgers University, 610 Taylor Rd., 
Piscataway, NJ 08854}
\author {Roberto Car}
\affiliation {Department of Chemistry and
Princeton Institute for the Science and
Technology of Materials (PRISM),
Princeton University, NJ 08544}
\author{Ralph Gebauer}
\affiliation {The Abdus Salam International Centre for Theoretical
Physics (ICTP), 34014 Trieste (Italy) \\
INFM/Democritos, National Simulation Center, 34013 Trieste, Italy}
\date{\today}

\begin{abstract}
Time-dependent density functional theory is extended to include
dissipative systems evolving under a master equation, providing
a Hamiltonian treatment for molecular electronics.
For weak electric
fields, the isothermal conductivity is shown to match the
adiabatic conductivity, thereby recovering the Landauer result.
\end{abstract}
\pacs{71.15.Mb, 31.10.+z,  03.65.Yz}

\maketitle

\newcommand{\upalda}{^{\scriptscriptstyle \rm ALDA}}
\newcommand{\downalda}{_{\scriptscriptstyle \rm ALDA}}
\newcommand{\vuc}{^{\scriptscriptstyle \rm VUC}}
\newcommand{\Tr}{{\rm Tr\; }}
\newcommand{\TrB}{{\rm Tr_{R}\; }}
\def\Trb#1{{\rm Tr\;}\left( #1 \right) }

Much recent interest has focused on
using single molecules as transistors for a new breed
of computers\cite{NR03}.   The complex nature of these
devices, especially the leads, suggests that their properties
can be sensitive to chemical details.  Thus we wish to
model the transport characteristics of
such devices with first-principles electronic
structure methods, such as density functional theory (DFT).
However, the traditional theorems of DFT do not apply
to systems carrying current in finite electric fields.

Valid applications of DFT
are derived from exact principles of quantum mechanics.
These involve proofs of
a one-to-one correspondence between densities and potentials,
i.e., that a given one-electron density can only be produced
by at most a single one-body potential, under a given set of
restrictions.
Ground-state DFT was established by Hohenberg and
Kohn\cite{HK64}, by proving that, for interacting electrons, a given
ground-state density can be produced by at most one ground-state
one-body potential.  We say the one-body potential, and hence all
other properties, is a {\em functional} of the ground-state density.

Similarly, Runge and Gross (RG) showed\cite{RG84}
that, for interacting electrons
in a given initial state, a given evolution of the one-body
density can be produced by at most one {\em time-dependent}
one-body potential, thus establishing the validity of time-dependent density
functional theory (TDDFT).  We usually start in a non-degenerate
ground-state, so that the initial-state dependence becomes a ground-state
density dependence, via Hohenberg-Kohn.  A slight variation on this, proven in
passing by RG, uses the one-body current density as
the basic variable, called time-dependent current-density functional
theory (TDCDFT).  This yields a more natural description of 
solids in uniform electric fields\cite{MSB03}, including the
 polarization-dependence
first identified by GGG\cite{GGG95}.

On the other hand, transport problems within traditional wavefunction
theory are usually handled within the Landauer formalism\cite{XDR02}.  The 
molecule and contacts are placed between two infinite reservoirs
at chemical potentials that differ by the voltage drop across the
molecule.  A standard integral over the Green's function and
coupling to the reservoirs, and Fermi occupation factors, then yields
the current, and thus the conductance.  This can be made exact by
using non-equilibrium Keldysh Green's functions\cite{SA04},
by imagining the system
begins in the distant past with reservoirs and molecule decoupled,
and the coupling turned on adiabatically to produce a steady-state current.
Unfortunately, 
as the decoupled initial state is not the ground state of
any Hamiltonian, the functional needed in TDCDFT differs in some
unknown way from the usual functional\cite{MBW02}.

The present state-of-the-art of DFT transport through single molecules
is embodied by 
calculations such as those of Refs. \cite{XDR02} and
\cite{DPLb00}, and many others.
A self-consistent Kohn-Sham (KS) calculation is performed
for a molecule trapped between two leads, and the
scattering states are inserted into the (two-terminal)
Landauer formula to calculate the conductance. 
At finite fields, the KS Green's function is significantly
distorted from its zero-field value.  As is well-recognized\cite{XDR02},
use of the ground-state KS Green's function in place of the
exact non-equilibrium Green's function is an unjustified approximation.
In particular, when coupling between the molecule and leads is
weak, the poles of this Green's function, representing
resonances of the molecule, are at the KS orbital energy differences, 
which are {\em not} the true excitations of
the system.  Acknowledging such limitations,
there have been several recent attempts to go beyond this
picture\cite{DG04,VT04}.

In the present paper, we prove a new density functional theorem
that encompasses transport at finite electric fields, by including
dissipation to phonons via a master equation.  We derive the
associated KS master equation.  The new exchange-correlation (XC)
potential reduces to that of TDCDFT in the limit of zero dissipation.
Finally, we show how all the ingredients for a realistic calculation
can be constructed.

We consider only symmetric leads.
To avoid the use of reservoirs, 
put the entire system on a long thin ring, and thread through the
center a solenoidal magnetic field.  This produces a spatially
uniform electric field throughout the entire system, and is
equivalent to a change of gauge\cite{K64}.  The system is finite, and nowhere
are there two different chemical potentials.
In the limit of zero bias, the Kubo response formalism
can be applied, in either traditional many-body theory or
DFT.  Using this geometry, Kohn and Kamenev\cite{KK01} showed how, by being careful
with the order of limits, one can, within time-dependent
Hartree theory, consider the response of the system to
an AC electric field of frequency $\omega$, and by taking
$\omega\to 0$, recover the two- and four-terminal
Landauer results.  This procedure has been
generalized to TDCDFT\cite{BGE04}.

However, an important difficulty arises when the field is
finite.  For a purely electronic system, the electrons will
accelerate indefinitely, and the current grow infinitely.
In reality, there is dissipation due to
scattering with phonons to bring the system to equilibrium.
This basic phenomenon is described 
by the quantum Liouville equation for the density
matrix of the entire
system of electrons and phonons, $S_{tot}$:
\ben
\frac{d S_{tot}(t)}{dt} 
= - i [ H_{tot}, S_{tot}(t) ]
\label{Stot}
\een
In the case of bulk transport, Kohn and Luttinger\cite{KL57} showed
how, for scattering from dilute impurities in weak fields, 
Eq. (\ref{Stot}) recovers the Boltzman equation, identifying
the diagonal elements of the electronic density matrix with
the distribution function.

In the quantum mechanics of dissipative systems, there is
a well-established procedure for incorporating the
effects of inelastic scattering with a reservoir into the
long-time evolution of the system.  In this case, the
total Hamiltonian consists of
the system Hamiltonian $H$
(the electrons) and a reservoir Hamiltonian
$R$ (the phonons), coupled by $K$.   The system
Hamiltonian contains $N$ interacting electrons.
The coupling is linear in the phonon-coordinates,
and involves only one-body forces on the system.
Before an initial time ($t=0$), both
electrons and phonons are in thermal equilibrium 
at temperature $T$.  
The exact system density matrix $S$ then 
satisfies 
\ben
\frac{d S(t)}{dt} 
= - i [ H(t), S(t) ] -i \Tr [K,S_{tot}(t)]
\label{S}
\een
where the
trace is over reservoir coordinates.   To derive 
a master equation for $S$ alone\cite{CDG92},
we coarse-grain over a time scale $\Delta t$ that
is long compared to electronic transitions and phonon
correlations, but short compared to the relaxation time, i.e.,
the time-scale on which the electrons are losing energy to
the reservoir.
This yields
\ben
\frac{d {\bar S}(t)}{dt} 
= - i [ H, {\bar S}(t) ] +{\cal C}[{\bar S}(t)]
\label{Sbar}
\een
where ${\cal C}$ is a superoperator, found by applying Fermi's Golden
rule to the scattering process, and determined by the coupling $K$ and
the reservoir spectral density. It is usually written in a basis of
eigenstates of the time-independent (many-body) Hamiltonian, $H | A
\rangle = E_A | A \rangle$. In this basis, ${\cal C} [{\bar S}]$ reads
\bea
{\cal C}[{\bar S}] = - \sum_{A,B}&& \Gamma_{A\to B} \left( L_{AB} L_{BA}
S + \right. \\
&& \left. S L_{AB} L_{BA} - 2 L_{BA} S L_{AB} \right),
\label{CS}
\eea
where the operators $L_{BA}$ represent a transition from state $A$ to
state $B$, and the transition probabilities $\Gamma_{A\to B}$ are
given by
\ben
\Gamma_{A\to B} = \left\{ 
\begin{array}{ll}
{\cal D}(\omega_{AB}) |\gamma_{AB}|^2 \left( {\bar n}(\omega_{AB}) + 1
\right) & E_A > E_B \\
{\cal D}(\omega_{BA}) |\gamma_{AB}|^2 {\bar n}(\omega_{BA})  & E_A <
E_B
\end{array}
\right.
\label{Gamma}
\een
in terms of the electron-phonon coupling elements $\gamma_{AB}$.
${\cal D}(\omega)$ is the density of states of phonons with frequency
$\omega$ ($\omega_{AB} = E_A - E_B$), and ${\bar n}(\omega) = 1/\left(
e^{\frac{\omega}{kT}} -1 \right)$ is the thermal occupation factor.
The transition probabilities satisfy detailed balance
\ben
\exp(- E_A/k_B T) \Gamma_{A\to B} = \exp(-E_B/k_B T) \Gamma_{B\to A},
\een
so that the steady-state solution of Eq. (\ref{Sbar}) yields the
thermal equilibrium density matrix of $H$.  No matter what the
initial density matrix, the steady-state solution is always the same.
Thus the master equation couples statistical mechanics to quantum
mechanics and, by including only secular contributions, goes beyond
the Schr\"odinger equation for the electrons to build in irreversible
evolution.   

As long as the coupling is weak, the rate
at which the system relaxes back to equilibrium
(the relaxation time) will be much longer than
all other scales, and the master equation applies.
While there is much discussion about the validity of
the master equation, and what physics is contained
within it, in what follows we take Eq. (\ref{Sbar}) as
given, with all its merits and flaws, and show how
to map the system to an effective single-particle
system.
Although the system Hamiltonian is time-independent,
the time-dependence in the master equation is generated by, e.g.,
starting in a non-equilibrium density matrix, which evolves
into the thermal equilibrium density matrix.  This involves
excitation and de-excitation of the electrons.

Our goal is describe a many-electron system, evolving
under a master equation, by a collection of non-interacting
electrons.  To do this, we allow the one-body potential of
$H$ to be time-dependent.   This breaks the connection between
${\cal C}$ and $H$, since the final $H$ might have an external
electric field turned on, but the initial $H$ not.
We then construct a Runge-Gross style proof that,
for fixed electron-electron interaction and
superoperator $\cal C$, and 
for a given initial density matrix $S_0$, no
two one-body potentials can give rise to the
same time-dependent density $n(\br t)$.
We assume that the potentials are Taylor-series
expandable around $t=0$, and that some coefficient in the
expansion is not uniform in space.  Begin with the equation of
motion for the current density:
\bea
\frac{d \langle \bj(\br) \rangle }{ dt}
&=& \Trb  {\bj(\br) \frac{d{\bar S}}{dt}}\nonumber\\
&=& -i \Trb {\bj(\br)[H,{\bar S}]}
+\Trb{\bj(\br){\cal C}[{\bar S}]}
\eea
where $\bj(\br) = \sum_i \delta(\br-\br_i) \bp_i + \bp_i \delta(\br-\br_i)$
is the current-density operator.
Using the fact that cyclic permutation does not alter the
trace, we can write the first term on the right as
$i\Trb {[H,\bj(\br)] S}$, which is the usual contribution
from evolution under a Hamiltonian.
Evaluating everything at $t=0$, and considering two 
systems with possibly different potentials but the same
initial density matrix and coupling, we find 
\ben
\frac{d\Delta \langle \bj(\br) \rangle }{ dt}\Big\vert_{t=0}
= -\n_0(\br) \Delta \nabla v\ext (\br t=0)
\een
just as in RG.  Thus two potentials that differ at $t=0$
give rise to two different currents.

One next shows that, for two systems whose initial
Hamiltonians are the same, and two operators that are identical
initially, but whose time-evolution differs,
\ben
\frac{d\Delta \langle A \rangle }{ dt}\Big\vert_{t=0}
=\Trb{\frac{\partial\Delta A }{ dt}\Big\vert_{t=0}
{\bar S}(0)}
\een
within the master equation,
because both the commutator and ${\cal C}$ are identical
in both systems at $t=0$.  Applying this result to the
equation of motion for the $k$-th derivative of the currents,
\begin{equation} \label{dkdtj}
\partial^{k+1}_0
\Delta{\bf j}(\bx) =
-n_{0}({\bf r})\; \nabla \partial^k_0
\Delta v\ext(\bx)  \quad,
\end{equation}
where $\partial^k_0 = (\partial^k/\partial t^k) \big|_{t=0}$.
Thus, any difference in any derivative of the potentials (other
than a constant), produces two different currents.  This establishes
a one-to-one correspondence between densities and currents.

We could stop here, as it is the TD current DFT that is needed\cite{BGE04}
for transport calculations.  But, for generality, we also
wish to establish a {\em density} functional theory,
the final step of the RG theorem uses continuity to show that
the densities must differ for two different potentials.  This 
follows in the master equation evolution, since
\ben
\frac{d \langle \n(\br) \rangle }{ dt}\Big\vert_{t=0}
= -\Trb {\nabla\bj(\br) {\bar S(0)}} + 
\Trb{\n(\br) {\cal C}[{\bar S}(0)]}
\label{cont}
\een
and the last term is the same in both systems.
The usual arguments about the vanishing of the potentials
sufficiently rapidly at large distances then suffice for finite
systems\cite{GK90}, or single-valuedness for periodic systems\cite{MSB03}.
Breakdown of continuity in the master equation occurs because,
in Eq. (\ref{cont}), the superoperator provides a correction to
the usual statement under Hamiltonian evolution, i.e., some momentum
is transferred to the reservoir.  But use of Eq. (\ref{cont}) restores
continuity, and the correction can even be written in terms of a
current\cite{GC04}.

We have established that the potential is a functional of the
time-dependent current density for a given interaction, statistics,
initial density matrix, and coupling.
In principle, we can  apply the same argument
with the interaction set to zero, to produce a set of time-dependent
KS equations whose one-body potential, $v\s[\n,S\s(0),{\cal C}](\br t)$,
is defined to yield the exact $\n(\br t)$, when evolved under the
master equation.   By subtracting the external potential and Hartree
contribution, we find an XC potential that has
the same dependencies as the KS potential, but also depends on the
initial density matrix of the interacting system.  Fortunately, we
can subsume all dependence on the initial density matrices into the
functional itself, by beginning in an equilibrium distribution for
both the interacting and non-interacting systems.  In that case,
the Mermin functional, which is just a functional of the initial density,
determines the initial density matrix. 

Constructed this way, the KS system has certain pathologies.
The superoperator in the many-body master equation is guaranteed
to vanish only on the many-body density matrix, not on the KS
density matrix.  To compensate for this, the corresponding KS
potential might need to evolve forever, even after the KS system has
settled into a steady state\cite{MBW02}.
Nevertheless, we can construct a practical KS scheme by constructing
a KS superoperator, ${\cal C}\s$.  To do so, we define $v\s(T)(\br)$ as
the KS potential in the Mermin functional at temperature $T$, i.e., the
potential that, when thermally occupied with non-interacting
electrons, reproduces the exact
one-electron density
at thermal equilibrium.  We then apply perturbation theory for a weak
interaction between non-interacting electrons in this potential and
the phonons in the reservoir, yielding the analogs to Eqs. (\ref{Gamma}) above:
\ben
\gamma_{i\to j} = \left\{ 
\begin{array}{ll}
{\cal D}(\omega_{ij}) |\gamma_{ij}|^2 \left( {\bar n}(\omega_{ij}) + 1
\right) & \epsilon_i > \epsilon_j \\
{\cal D}(\omega_{ji}) |\gamma_{ij}|^2 {\bar n}(\omega_{ji})  & \epsilon_i <
\epsilon_j
\end{array}
\right.
\label{GammaKS}
\een
where $\epsilon_i$ are the eigenvalues of $-\nabla^2/2+v\s(T)(\br)$. 
The matrix elements $\gamma_{ij}$ are now evaluated for the interaction
between the KS system and the bath.
To find the KS master equation itself, we reduce the many-body
Eq.~(\ref{Sbar}) to a single particle form by tracing out all other
degrees of freedom, and using a Hartree-style approximation for the
two-particle correlation functions appearing in ${\cal C}[{\bar
 S}]$. In the basis of the single-particle KS orbitals, $|n \rangle,
|m \rangle, |p \rangle$, we find\cite{GCb04}:
\bea
\frac{d s_{nm}}{dt} &=& -i \sum_p \left( h_{np} s_{pm} - s_{np} h_{pm}
\right) \nonumber \\
&& + \left( \delta_{nm} - s_{nm}\right) \sum_p \left( \gamma_{np} +
\gamma_{mp} \right) s_{pp} \\
&& - s_{nm} \sum_p  \left( \gamma_{pn} + \gamma_{pm} \right) \left(
1 - s_{pp}\right). \nonumber
\eea
In this KS master equation, the steady-state equilibrium has a
static potential.  The approach rate, determined by $\gamma$, will
not match that of the true system, but such effects are absorbed
in the XC potential.  The important point is that, if such a KS
system exists, it is unique for the given coupling, by the theorem
proven above.

Next we discuss the XC functional, which depends on the coupling
to the reservoir.  One could imagine performing accurate
wavefunction calculations for a uniform gas on a ring, with the
given coupling, to produce a local density approximation for the master
equation.  But we argue that the usual approximations of
TDDFT (or TDCDFT, as needed), such as the adiabatic local
density approximation, are likely
to suffice, because the KS master equation includes dissipation,
and drives the system to the thermodynamic steady state.
Thus the effect of dissipation on XC is likely
to be small, and might even vanish in the limit of weak coupling to
the reservoir. 
The important XC effects are to correct the
electronic transition frequencies into the true transitions, and
this is captured exactly by such an approximation.

The dissipative part of the master equation, ${\cal C}[{\bar S}]$ in
Eqs.~(\ref{Sbar}) and (\ref{CS}) depends on phonon frequencies
$\omega$, densities of state ${\cal D}(\omega)$, and coupling matrix
elements $\gamma_{nm}$. All these quantities can be extracted from
first-principles density-functional linear response
calculations\cite{BGCG01}. This allows for a fully consistent DFT
implementation of the dissipative dynamics.

Lastly, we discuss the recovery of the Landauer result for
the case of transport through a single molecule, in the limit
of weak bias.  This has recently\cite{BGE04} been derived using
TDCDFT.  Here we show the derivation of linear response for the
more familiar DFT, and the current version
is essentially the same (but more cumbersome).

For a given KS master equation, assume the density matrix has
evolved into its steady state, so that
\ben
[H_0,\bar S_0]= - i {\cal C}\s({\bar S_0})
\label{ss}
\een
If ${\cal C}\s$ has been constructed to thermalize the eigenstates
of $H_0$, both sides of this equation vanish.
Now imagine perturbing the system with a weak time-dependent
potential $\Delta V$, which becomes constant
after a finite time.  Allow the system to relax back to 
its new steady state, with density matrix ${\bar S}_0 + \Delta S$.
Equating equal
powers in the perturbation:
\ben
[H_0,\Delta S] +i \eta[{\bar S_0}]\cdot{\Delta S}= -[\Delta V, {\bar S}_0] 
\een
where $\eta$ is the first derivative of ${\cal C}\s$.
Expanding all quantities in eigenstates of $H_0$ and solving, we find
\ben
\Delta S_{AB} = \frac{f_A V_{AB} - f_B V_{AB}}{E_A-E_B + i \eta_{AB}}
\label{SAB}
\een
where $\bar S_0=\sum |A\rangle f_A \langle A|$, i.e., $f_A$ is the
Fermi occupation factor of state $A$.  Calculating the density
change by tracing the density operator with $\Delta S$, and
recognizing that the system eigenstates are Slater determinants
of orbitals, one recovers the usual 
density-density response function\cite{GK85}.
A similar derivation holds for the current-current response
within TDCDFT.

To understand what this means, consider
a KS master equation with weak 
dissipation ($\eta_{ij} << \omega_{ij}$), on an infinite
ring.  Turn on a small but
finite electric field, and evolve the system into a steady
state.  This is the isothermal conductivity\cite{KL57}, found from the
steady-state solution for the electrons coupled to the phonons.
The derivation above shows that this reduces to the adiabatic
conductivity as given by the Kubo response formula, which
has 
recently been shown to
recover the Landauer result (with possible XC corrections)\cite{BGE04}.

Acknowledgments:  We gratefully acknowledge financial support from
DOE under grant number DE-FG02-01ER45928, 
stimulating discussions with Walter
Kohn, Tchavdar Todorov, and Stefano Sanvito.  Some of this work
was performed at the Centre for Research in Adaptive Nanosystems
(CRANN) supported by Science Foundation Ireland (Award 5AA/G20041).

\end{document}